**Bibliometrics/Citation networks**    (preprint version)
in: George A. Barnett (Ed.), The Encyclopedia of Social Networks, Sage, 2011, pp. 72-74.

In addition to shaping social networks, for example, in terms of co-authorship relations, scientific communications induce and reproduce cognitive structures. Scientific literature is intellectually organized in terms of disciplines and specialties; these structures are reproduced and networked reflexively by making references to the authors, concepts and texts embedded in these literatures. The concept of a cognitive structure was introduced in social network analysis (SNA) in 1987 by David Krackhardt, but the focus in SNA has hitherto been on cognition as a psychological attribute of human agency. In bibliometrics, and in science and technology studies (STS) more generally, socio-cognitive structures refer to intellectual organization at the supra-individual level. This intellectual organization emerges and is reproduced by the collectives of authors who are organized not only in terms of inter-personal relations, but also more abstractly in terms of codes of communication that are field-specific. Citations can serve as indicators of this codification process.

Citation indexing has a long tradition: Bella Weinberg mentions that the first citation index for the Talmud was printed in Italy between 1522 and 1524. However, the modern citation (with its standardized format) and the modern citation index can be considered as textual innovations which enable scholars to communicate across different literatures. The science citation indices were shaped by Eugene Garfield at the Institute of Scientific Information in the 1960s and 70s. In 1965, the historian of science Derek de Solla Price wrote a foundational article for this field of study entitled "Networks of scientific papers." Price proposed to study, among other things, the preferential attachment mechanism among scientific papers in terms of a negatively exponential function (in STS also known as Robert Merton's [1968] "Matthew effect"), aggregated journal-journal citation networks as an operationalization of specialty structures, and the dynamics of articles and reviews in terms of variation, selection and retention mechanisms.

Bibliometrics adds to SNA a focus on content. SNA methods and techniques pervaded the bibliometric domain after the emergence of the internet during the 1990s and the increased attention to network dynamics in various disciplines. Scientific communication, however, abstracts to a large extent from the historical carriers of communications in favor of the intellectual organization in the constructs. Paradoxically, the constructed takes precedence over the constructors, and social relations among the latter tend to be reconstructed accordingly. While social relations play a role in the bottom-up dynamics of network construction, from a top-down perspective their role changes to that of a potentially dependent variable. Citations and other indicators therefore can be expected to have different meanings in the social and/or intellectual organization of the sciences. Despite the continuous call for a single theory of citations, these systems of reference provide different contexts of relevance.

In the social domain, for example, numbers of citations and publications are increasingly used for measuring scientific performance in ranking exercises. Citedness can also be considered as the aggregate of weighted indegree. However, most scientometric indicators did not originate from SNA. Indicators used for ranking and evaluation are based on vectors and not on matrices.

Mapping efforts presume an underlying matrix representation. Citation maps and dynamic animations of bibliometric data enable us to position and visually trace new developments in the sciences in terms of the emergence of new (and potentially interdisciplinary) fields, bursts of activities, bifurcations, and mergers. Maps can also be used to evaluate the effects of science policy interventions. Visualizations, however, tend to be user-oriented and therefore flexible. The pragmatics of usefulness are traded off against analytical objectives. From an analytical perspective, visualization can be helpful to the interpretation of algorithmically warranted results.

As noted, bibliometric data are usually weighted. In graph theory which provides one of the theoretical bases for SNA, weighting is a next step, whereas the focus is on proving and elaborating algorithms first for the unweighted graph. Furthermore, from this perspective the 1-mode matrix has priority, while in bibliometrics the focus is on document-word or document-author matrices which can be considered as 2-mode or, more traditionally, attribute matrices. The current tendency in social network analysis to pay increasing attention to 2-mode matrices may further bridge this gap between the two traditions. However, the differences in starting points (document sets versus network relations) are analytical and can thus be expected to remain relevant.

One important consequence of this difference in perspectives is the potentially different topologies in the representations. Graph-theoretical analysis focuses on the network that can be observed in terms of relations among agents. However, the relations span a network with an architecture in which agents also have a position. The latent dimensions of this multi-dimensional space can be analyzed using techniques such as factor analysis and multi-dimensional scaling. (Note that these techniques were developed for attribute matrices and not for 1-mode affiliations matrices.) The latent dimensions span a vector space which can be projected on a map for the purposes of the visualization of positions. Relations can then be added, but from a starting point different from a spring-embedded algorithm that assumes relations as its basic material.

I distinguish the two perspectives because mixing these analytical starting points may lead to confusion. For example, in the otherwise impressive tradition of developing author co-citation analysis (ACA), the 1-mode affiliation matrices are sometimes factor-analyzed when one could have used the underlying 2-mode attribute matrices. In the meantime, ACA has become a standard technique in citation analysis, based on the co-citation of two authors (or documents) by a third author (or publication) which cites both of them. Since this operation is recursive, one can consider networks of citations as developing over time. This dynamic perspective can be elaborated into an algorithmic historiography of science, for which nowadays dedicated software is also available. Algorithms from social network analysis such as main-path analysis can further elucidate the structure. Thus, combinations of SNA with bibliometrics are increasingly common.

The concept of co-citation analysis can be generalized, for example, to journal co-citation analysis. Aggregated journal-journal citation matrices have been used to map the sciences in terms of journals, specialties, and disciplines. Journals in similar fields can be expected to cite one another in networks more dense than across fields. However, interdisciplinary (at the level of two disciplines such as *Limnology and Oceanography*) and transdisciplinary journals (such as *Science* and *Nature*) also play important roles in scholarly communication to the extent that a publication in one of these journals may not only illustrate the "strength of weak ties," but also integrate at a next-order level what tends to be continuously differentiated in terms of the codified jargons of specialties.

Research fronts are further developed in the parallel processes at the specialty level. Variation is introduced into the publication system by new knowledge claims in submitted materials. Selection operates recursively on this variation. First, each paper is positioned (by authors, editors, and reviewers) amidst other papers in a network of scientific communications. The author may have made references to some of these, but a reader may intuit links with other literature. Selecting the contribution as one of the references in a next paper reconstructs the position of the cited paper as relevant to further knowledge claims. This selection over time on the selections implied in positioning the paper at the time of publication can stabilize (or destabilize) its knowledge claim and make it increasingly part of a cognitive structure, while previously it was only a possible point of reference. When the paper becomes incorporated into the scientific literature, the reference may be obliterated or it may become a citation classic. This "obliteration by incorporation" can also be considered as symbolic generalization, that is, the knowledge becomes part of a latent code which steers the communication at the disciplinary level of the field.

This model of recursive loops operating as feedbacks upon networked fluxes of communications can be formalized with the model of *autopoiesis* or self-organization that has been studied, for example, in biology. However, when molecules self-organize life as an emergent phenomenon, the subject matter remains tangible. The self-organization of the meaning of textual messages into discursive knowledge by interacting readings among scholars sets a research agenda at the interface between "evolutionary" bibliometrics, SNA, and STS. Knowledge, for example, can be considered as a meaning that makes a difference with reference to a code of communication emerging in a network. In a knowledge-based economy, the networks of knowledge communication become increasingly a third social coordination mechanism in addition to the economic forces of the market and normative exchanges in policy-making.

See also: Innovation Networks, Semantic Networks, Self-organizing Networks

Further reading:
Garfield, E. *Citation Indexing: Its Theory and Application in Science, Technology, and Humanities*. New York: John Wiley, 1979.
Hummon, N.P., & Doreian, P. Connectivity in a citation network: The development of DNA theory. *Social Networks*, 11, 39-63 (1989).
Krackhardt, D. Cognitive Social Structures, *Social Networks* 9, 109-134 (1987).

Leydesdorff, L. Scientific Communication and Cognitive Codification: Social Systems Theory and the Sociology of Scientific Knowledge, *European Journal of Social Theory,* 10(3), 375-388 (2007).
Leydesdorff, L. The Knowledge-Based Economy and the Triple Helix Model. *Annual Review of Information Science and Technology,* 44, 367-417 (2010).
Luhmann, N. *Social Systems*. Stanford, CA: Stanford University Press, 1995.
Merton, R. K. The Matthew Effect in Science. *Science,* 159, 56-63 (1968)
Moed, H. F., Glänzel, W., & Schmoch, U. *Handbook of Quantitative Science and Technology Research: The Use of Publication and Patent Statistics in Studies of S & T Systems*. Dordrecht, etc.: Springer, 2004.
Price, D. J. de Solla, Networks of scientific papers. *Science,* 149, 510- 515 (1965).
Weinberg, B. H. The Earliest Hebrew Citation Indexes, *Journal of the American Society for Information Science* 48(3), 318-330 (1997).

Loet Leydesdorff
Amsterdam School of Communications Research (ASCoR),
University of Amsterdam.